\documentclass[twocolumn,runningheads]{svjour2}
\smartqed  % flush right qed marks, e.g. at end of proof
\usepackage{graphicx}
\journalname{Astrophysics and Space Science}
\begin{document}

\title{Underground Water Cherenkov Muon Detector Array \\ with the Tibet Air Shower Array \\ for Gamma-Ray Astronomy in the 100 TeV Region
}
\subtitle{}

\titlerunning{Underground Water Cherenkov Muon Detector Array}        % if too long for running head

\author{
M.~Amenomori \and S.~Ayabe \and X.J.~Bi \and D.~Chen \and S.W.~Cui \and Danzengluobu \and
L.K.~Ding \and X.H.~Ding \and C.F.~Feng \and Zhaoyang~Feng \and Z.Y.~Feng \and X.Y.~Gao \and Q.X.~Geng \and
H.W.~Guo \and H.H.~He \and M.~He \and K.~Hibino \and N.~Hotta \and Haibing~Hu \and H.B.~Hu \and
J.~Huang \and Q.~Huang \and H.Y.~Jia \and F.~Kajino \and K.~Kasahara \and Y.~Katayose \and
C.~Kato \and K.~Kawata \and Labaciren \and G.M.~Le \and A.F.~Li \and J.Y.~Li \and H.~Lu \and S.L.~Lu \and
X.R.~Meng \and K.~Mizutani \and J.~Mu \and K.~Munakata \and A.~Nagai \and H.~Nanjo \and
M.~Nishizawa \and M.~Ohnishi \and I.~Ohta \and H.~Onuma \and T.~Ouchi \and S.~Ozawa \and
J.R.~Ren \and T.~Saito \and T.Y.~Saito \and M.~Sakata \and T.K.~Sako \and T.~Sasaki \and
M.~Shibata \and A.~Shiomi \and T.~Shirai \and H.~Sugimoto \and M.~Takita \and Y.H.~Tan \and
N.~Tateyama \and S.~Torii \and H.~Tsuchiya \and S.~Udo \and B.~Wang \and H.~Wang \and X.~Wang \and Y.G.~Wang \and
H.R.~Wu \and L.~Xue \and Y.~Yamamoto \and C.T.~Yan \and X.C.~Yang \and S.~Yasue \and Z.H.~Ye \and
G.C.~Yu \and A.F.~Yuan \and T.~Yuda \and H.M.~Zhang \and J.L.~Zhang \and N.J.~Zhang \and
X.Y.~Zhang \and Y.~Zhang \and Yi~Zhang \and Zhaxisangzhu \and X.X.~Zhou (The Tibet AS$\gamma$ Collaboration)
}

\authorrunning{M.~Amenomori et al.} % if too long for running head

\institute{
M.~Amenomori \and H.~Nanjo \at
Department of Physics, Hirosaki University, Hirosaki 036-8561, Japan
\and
S.~Ayabe \and K.~Mizutani \and H.~Onuma \at
Department of Physics, Saitama University, Saitama 338-8570, Japan
\and
X.J.~Bi \and L.K.~Ding \and Zhaoyang~Feng \and H.H.~He \and H.B.~Hu \and H.~Lu \and S.L.~Lu \and J.R.~Ren \and Y.H.~Tan \and H.~Wang \and H.R.~Wu \and H.M.~Zhang \and J.L.~Zhang \and Y.~Zhang \and Yi~Zhang \at
Key Laboratory of Particle Astrophysics, Institute of High Energy Physics, Chinese Academy of Sciences, Beijing 100049, China
\and
D.~Chen \and Y.~Katayose \and M.~Shibata \at
Faculty of Engineering, Yokohama National University, Yokohama 240-8501, Japan
\and
S.W.~Cui \at
Department of Physics, Hebei Normal University, Shijiazhuang 050016 , China 
\and
Danzengluobu \and X.H.~Ding \and H.W.~Guo \and Haibing~Hu \and Labaciren \and X.R.~Meng \and A.F.~Yuan \and Zhaxisangzhu \at
Department of Mathematics and Physics, Tibet University, Lhasa 850000, China
\and
C.F.~Feng \and M.~He \and A.F.~Li \and J.Y.~Li \and Y.G.~Wang \and L.~Xue \and N.J.~Zhang \and X.Y.~Zhang \at
Department of Physics, Shandong University, Jinan 250100, China
\and
Z.Y.~Feng \and Q.~Huang \and H.Y.~Jia \and G.C.~Yu \and X.X.~Zhou \at
Institute of Modern Physics, South West Jiaotong University, Chengdu 610031, China
\and
X.Y.~Gao \and Q.X.~Geng \and J.~Mu \and B.~Wang \and X.C.~Yang \at
Department of Physics, Yunnan University, Kunming 650091, China
\and
K.~Hibino \and T.~Ouchi \and T.~Sasaki \and T.~Shirai \and N.~Tateyama \and T.~Yuda \at
Faculty of Engineering, Kanagawa University, Yokohama 221-8686, Japan
\and
N.~Hotta \at
Faculty of Education, Utsunomiya University, Utsunomiya 321-8505, Japan
\and
J.~Huang \and K.~Kawata \and M.~Ohnishi \and S.~Ozawa \and T.Y.~Saito \and T.K.~Sako \and A.~Shiomi \and M.~Takita \and S.~Udo \and X.~Wang \and C.T.~Yan \at
Institute for Cosmic Ray Research, University of Tokyo, Kashiwa 277-8582, Japan 
\\\email{kawata@icrr.u-tokyo.ac.jp}
\and
F.~Kajino \and M.~Sakata \and Y.~Yamamoto \at
Department of Physics, Konan University, Kobe 658-8501, Japan
\and
K.~Kasahara \at
Faculty of Systems Engineering, Shibaura Institute of Technology, Saitama 337-8570, Japan
\and
C.~Kato \and K.~Munakata \at
Department of Physics, Shinshu University, Matsumoto 390-8621, Japan
\and
G.M.~Le \and Z.H.~Ye \at
Center of Space Science and Application Research, Chinese Academy of Sciences, Beijing 100080, China
\and
K.~Mizutani \and S.~Torii \at
Advanced Research Institute for Science and Engineering, Waseda University, Tokyo 169-8555, Japan
\and
A.~Nagai \at
Advanced Media Network Center, Utsunomiya University, Utsunomiya 321-8585, Japan
\and
M.~Nishizawa \at
National Institute of Informatics, Tokyo 101-8430, Japan
\and
I.~Ohta \at
Tochigi Study Center, University of the Air, Utsunomiya 321-0943, Japan
\and
T.~Saito \at
Tokyo Metropolitan College of Industrial Technology, Tokyo 116-8523, Japan
\and
H.~Sugimoto \at
Shonan Institute of Technology, Fujisawa 251-8511, Japan
\and
H.~Tsuchiya \at
RIKEN, Wako 351-0198, Japan
\and
S.~Yasue \at
School of General Education, Shinshu University, Matsumoto 390-8621, Japan 
}

\date{Received: date / Accepted: date}

\maketitle

\begin{abstract} 

 We propose to build a large water-Cherenkov-type muon-detector array
(Tibet MD array) around the 37,000~m$^{2}$ Tibet air shower array
(Tibet AS array) already constructed at 4,300~m above sea level in
Tibet, China.  
Each muon detector is a waterproof concrete pool, 6~m
wide $\times$ 6~m long $\times$ 1.5~m deep in size, equipped with a 20
inch-in-diameter PMT.  
The Tibet MD array consists of 240 muon
detectors set up 2.5~m underground. Its total effective area will be
8,640 m$^{2}$ for muon detection.  The Tibet MD array will
significantly improve gamma-ray sensitivity of the Tibet AS array in
the 100~TeV region (10-1000~TeV) by means of gamma/hadron separation
based on counting the number of muons accompanying an air shower.
The Tibet AS+MD array will have the sensitivity to gamma rays in the 100 TeV region by an order of magnitude
better than any other previous existing detectors in the world.

\keywords{Gamma ray \and Cosmic ray \and Muon \and SNR}
\PACS{95.55.Ka \and 98.70.Sa \and 95.85.Ry}
\end{abstract}

\section{Introduction}
\label{intro}
 Based on observation by 4 large imaging air Cherenkov telescopes in
Namibia, the HESS group recently reported the discovery of 14 new
gamma-ray sources in the TeV region, among which 8 sources are
UNIDentified (UNID) or SuperNova Remnant (SNR)-like \cite{Aha06}. Many
of the 14 sources have a harder energy spectrum (indices: $-$1.8 to
$-$2.8) at TeV energies than the standard candle Crab (index: $-$2.6).
These energy spectra were measured from 200~GeV and turned out to
extend up to 10~TeV approximately.  Cosmic rays are supposed to be
accelerated up to the knee energy region at SNRs in our
galaxy. Therefore, we naturally expect gamma rays in the 100~TeV
region (10-1000~TeV) which originate in $\pi^{0}$ decays produced by
the accelerated charged cosmic rays interacting with matter
surrounding the SNRs. 

Here, we will demonstrate our excellent sensitivity of
the Tibet air shower array plus muon detector array (Tibet AS+MD
array) to gamma rays in the 100~TeV region calculated by the Monte Carlo (MC) simulation
and give some speculation on 100~TeV gamma-ray source candidates.

\section{The Tibet AS+MD array}
\label{sec:1}

\begin{figure}
\centering
  \includegraphics[width=0.35\textwidth]{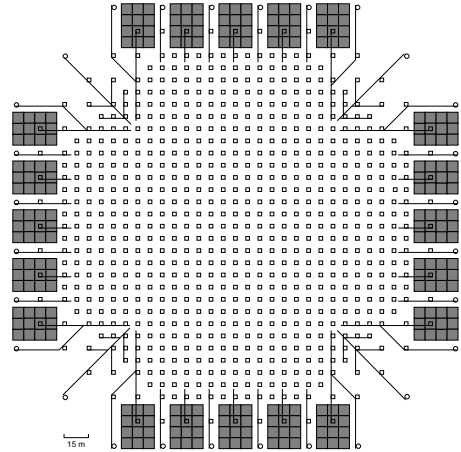}
\caption{
Schematic view of the Tibet AS+MD array. Open squares and open circles
show scintillation detectors which compose the Tibet AS array in
operation now.  Filled squares show the proposed water Cherenkov muon
detector array (Tibet MD array) buried 2.5~m underground.
}
\label{fig:1}       % Give a unique label
\end{figure}

 The Tibet air shower experiment has been successfully operated at
Yangbajing (90.522$^{\circ}$~E, 30.102$^{\circ}$~N, 4,300~m above sea
level) in Tibet, China since 1990.  
At present, the Tibet AS array consists of 789 plastic
scintillation counters of 0.5~m$^2$ each viewed by the 2
inch-in-diameter photo-multiplier tube (PMT), placed on a 7.5~m square
grid with an enclosed area of 37,000~m$^2$ to detect high-energy ($>$
a few TeV) cosmic-ray air showers as shown by open squares and circles in
Figure~\ref{fig:1} \cite{Ame05c}.  

 We are planning to build a water-Cherenkov-type muon detector
array (Tibet MD array) around the Tibet AS array.  Each muon detector is a waterproof
concrete pool, 6~m wide $\times$ 6~m long $\times$ 1.5~m deep in size, equipped
with a 20 inch-in-diameter (20$''\phi$) PMT (HAMAMATSU R3600).  
The inside of the concrete pool is painted with white
epoxy resin to waterproof and to efficiently gather catoptric water
Cherenkov lights by a downward facing PMT.  The Tibet MD array
consists of 240 muon detectors set up 2.5~m underground (2.0~m soil +
0.5~m concrete ceiling: $\sim$19 radiation length) as shown by gray areas in
Figure~\ref{fig:1}. Its total effective area amounts to be 8,640~m$^{2}$ for
muon detection with an energy threshold of 1~GeV.  
This configuration of the Tibet MD array is simply one example for
easy installation under the existing Tibet AS array.
Advantages of the
underground water Cherenkov detector are high cost performance and high sensitivity to muons
rather than the electromagnetic component caused by the environmental background
radioactivity and the air shower cascade, because it is easy to design
its pool depth (= path length of a muon) deeper, compared with a scintillation
detector.
 
 The dark pulse rate of 20$''\phi$ PMT measured in laboratory is
typically 100~Hz above 10 photoelectrons (PEs) at 
10$^{7}$ gain at 1500~V supplied voltage.  In this case, 
relativistic muons made by primary cosmic rays near the top of
atmosphere dominate the accidental background at approximately
10~kHz per 36~m$^{2}$ muon detector.  
Our DAQ system will enable us to set the time window for
muons to 200~ns by the offline analysis, therefore, the accidental
event rate will be 0.5 ($10100~{\rm Hz} \times 240 \times 200~{\rm ns}$) events per air shower trigger.
The water for the Tibet MD array ($\sim$13~kton) will be supplied from abundant
underground water pumped up at a village adjacent to our site.  According to an
examination of this water, the light attenuation length is longer than
several tens of meters.  The water in the pools will be continuously circulated
through 0.1 $\mu$m mesh filter and UV sterilizer to keep the light
attenuation length.  The water never freezes and bacteria do not proliferate
easily, since the temperature at 2.5~m underground remains stable and
cold between 3$^{\circ}$C and 12$^{\circ}$C through the year.

\section{MC simulation}

 The air shower events induced by cosmic rays and gamma rays are
generated by the Corsika Ver.6.204 code \cite{Cor98} with QGSJET01c for
the hadronic interaction model.  Primary cosmic rays are thrown along
Crab's orbit around the Earth, and the relative chemical composition
model \cite{Ame06} of primary cosmic rays is adopted based mainly on direct
observational data in the energy range from 0.3~TeV to 10000~TeV.
Primary gamma rays are also thrown along Crab's orbit
around the Earth assuming a differential power-law spectrum of
$E^{-2.6}$ in the energy range from 0.3~TeV to 10000~TeV.  Air shower
events are uniformly thrown within a circle radius 300~m centered at
the array center. This radius is sufficient to collect all air shower
events which are actually triggered in our array.

 The simulation of the Tibet AS array including the scintillation
detector response was already established based on the Epics uv8.00
code \cite{Epics}.  Finally, we get the estimated/true air shower
direction, core position, the sum of particle density for all detectors
($\Sigma\rho$) related to true energy, and so on for each air shower event.
Distributions of the MC events, such as parameters mentioned above, etc, are consistent with experimental data.
The trigger condition is imposed by the Tibet AS array, i.e., each shower
event should fire four or more of the scintillation detectors recording
1.25 particles or more. The energies at 100\% trigger
efficiency are estimated to be approximately 10~TeV for gamma rays and
30~TeV for cosmic rays, respectively.  After the air shower
reconstruction and some event selections, the angular resolution and
energy resolution are also estimated to be approximately
0.2$^{\circ}$ and 40\% at 100~TeV, respectively.

 The response of the water Cherenkov muon detector and the soil as an
absorber are simulated based on GEANT4~8.0 code \cite{Gea03},
considering the detailed structure of the Tibet MD array.
First, we trace the secondary particles of air showers triggered by
the Tibet AS array in the underground soil (2.0~g/cm$^{3}$: 70\% SiO$_{2}$,
20\% Al$_{2}$O$_{3}$, 10\% CaO).  All surviving particles under the
soil with energies exceeding the Cherenkov threshold are subsequently
fed into the simulation of the Tibet MD array including 0.5~m thick concrete
(2.3 g/cm$^{3}$: 100\% SiO$_{2}$) ceilings and walls.  The reflectance
at the surface of walls is assumed to be 70\% with isotropic
reflection.  After we simulate Cherenkov radiation, propagation of
Cherenkov photons in water, and the response of 20$''\phi$ PMT, the
number of photoelectrons ($N_{\rm PE}$) is counted up for each muon
detector.  The quantum efficiency of 20$''\phi$ PMT for wavelengths of
340-400~nm is approximately 20\%.  The peak of $N_{\rm PE}$ 
is estimated to be 26~PEs with width $\sigma_{N_{\rm PE}}$ +130\%$-$30\% approximately for one muon detector
when a vertical muon passes through a muon detector.

\section{Results and discussions} 

\begin{figure}
\centering
  \includegraphics[width=0.50\textwidth]{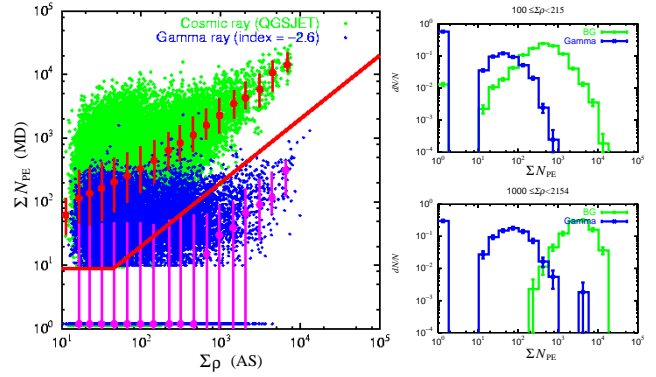}
\caption{ 
Distribution of $\Sigma N_{\rm PE}$ for the Tibet MD array as a
function of $\Sigma \rho$ for the Tibet AS array.
Blue and green points indicate gamma-induced and
hadron-induced air showers, respectively.  Closed circles and error bars show
median, 80\% and 20\% values, respectively. The cut to suppress
hadron-induced air showers is shown as a solid line.  Air showers with
no recorded PE by the Tibet MD array are plotted as $\Sigma N_{\rm
PE}$ = 1.2. Upper right: $\Sigma N_{\rm PE}$ distribution in typical energy band 10 TeV
(100$\leq$$\Sigma\rho$$<$215). Lower right: 100 TeV (1000$\leq$$\Sigma\rho$$<$2154).
}
\label{fig:2}       % Give a unique label
\end{figure}

\begin{figure}
\centering
  \includegraphics[width=0.45\textwidth]{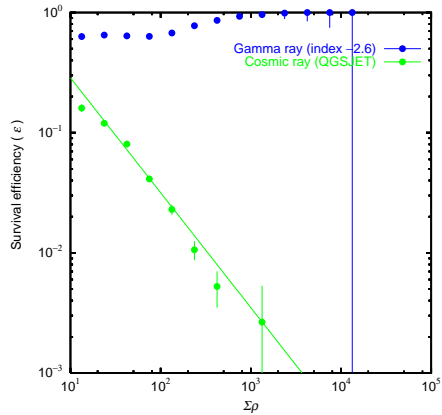}
\caption{
Survival efficiency after the cut.  Blue and green 
circles show gamma-induced and hadron-induced air showers,
respectively.
}
\label{fig:3}       % Give a unique label
\end{figure}

\begin{figure}
\centering
  \includegraphics[width=0.4\textwidth]{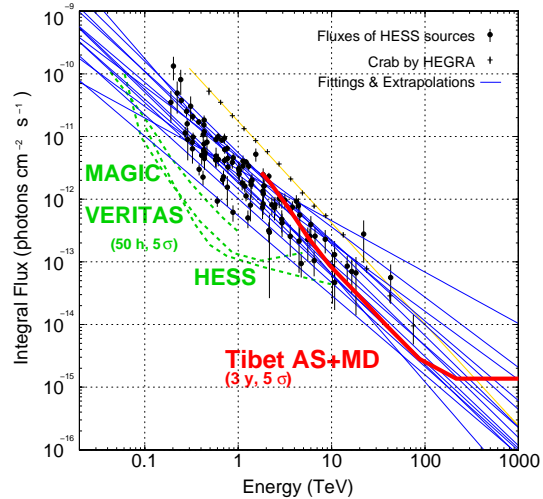} 
\caption{ Integral flux sensitivities to point-like gamma-ray sources. Dashed curves show
sensitivities of Cherenkov telescopes at 5 $\sigma$ for 50 hours
(MAGIC, VERITAS and HESS from the upper curve). The thick solid curve demonstrates the sensitivity
of the Tibet AS+MD array at 5 $\sigma$ for 3 calendar years.
Closed circles show the integral fluxes converted from
observed differential fluxes of ``HESS J'' sources point by point assuming
their spectral indices \cite{Aha06}. Thin lines show fittings and
extrapolations to HESS data points. } 
\label{fig:4} % Give a unique label 
\end{figure}

Figure~\ref{fig:2} shows the distribution of $\Sigma N_{\rm
PE}$ for the Tibet MD array as a function of $\Sigma \rho$ for the
Tibet AS array by the MC simulation, where $\Sigma N_{\rm PE}$ denotes
the sum of $N_{\rm PE}$ for muon detectors fired with a threshold
$N_{\rm PE} > 10$ 
and $\Sigma \rho$
is the sum of particle density for all scintillation detectors,
$\Sigma \rho$ = 1000 corresponds approximately to 100~TeV primary
cosmic-ray energy.  $\Sigma N_{\rm PE}$ also takes the accidental
muons into account by Poisson distribution with an average 0.5 muons/event as
described in $\S$\ref{sec:1}.  To select muon-poor air showers, we
optimize the cut condition on $\Sigma N_{\rm PE}$-$\Sigma \rho$ plot
as shown by the solid line in Figure~\ref{fig:2}.  Figure~\ref{fig:3} shows the
survival efficiency after the cuts by the MC simulation. Hadron-induced air
showers are suppressed by 99\% around $\Sigma \rho$ = 1000,
while gamma ray-induced air showers remain by more
than 95\%. Finally, we calculate the integral flux sensitivity of the
Tibet AS+MD array to gamma rays as shown by the thick solid curve in
Figure~\ref{fig:4}. Note that our sensitivity above 200~TeV is defined as
a flux corresponding to 15 gamma-ray events, since the background
events are fully suppressed to less than one event.

 Then, how many known/unknown sources do we expect to detect by the
Tibet AS+MD array, assuming the energy spectra of the gamma-ray
sources extended up to the 100 TeV region? The diffuse gamma rays from
the Cygnus region reported by the Milagro group \cite{Atk05},
TeV~J2032+4130 \cite{Aha05}, HESS~J1837-069 \cite{Aha06}, Crab, Mrk421
are clearly detectable. Cas~A \cite{Aha01}, HESS~J1834-087
\cite{Aha06}, and M87 \cite{Bei04} are marginal. In addition to these
existent/established sources, we can expect unknown sources in the
northern sky. Most of the HESS 14 sources would be detected by the
Tibet AS+MD array, as is shown in Figure~\ref{fig:4}, if it were located
at the HESS site. There exist approximately 80 SNRs within the scanned
area by the HESS telescopes, while the Tibet AS+MD array also observes
80 SNRs within its field-of-view. This in turn means that we can
expect to discover a dozen new sources, half of which will be UNID
or SNR-like, in the northern sky where no extensive search has been
done by an apparatus with sensitivity comparable to HESS.
The MAGIC and VERITAS experiments in
the northern hemisphere, together with the Tibet AS+MD array will
contribute to a deeper understanding of the origin and  acceleration mechanism
of cosmic rays.

%\begin{figure*}
%\centering
%  \includegraphics[width=0.75\textwidth]{fig1.eps}
%\end{figure*}
%

\begin{acknowledgements}
This work is supported in part by Gra\-nts-in-Aid for Scientific
Research and by Scientific Research (JSPS) in
Japan, and by the Committee of the Natural Science Foundation and by
Chinese Academy of Sciences in China.
\end{acknowledgements}

% BibTeX users please use
%\bibliographystyle{spmpsci}


\end{document}